\documentclass[aps,twocolumn,showpacs]{revtex4}
\usepackage{graphics}
\usepackage{graphicx}

\begin{document}
\title{Carrier-induced modulation of radiation by a gated graphene}
\author{M.V. Strikha}
\author{F.T. Vasko}
\email{ftvasko@yahoo.com}
\affiliation{Institute of Semiconductor Physics, NAS of Ukraine, Pr. 
Nauky 41, Kyiv, 03028, Ukraine}
\date{\today}

\begin{abstract}
The modulation of the transmitted (reflected) radiation due to change 
of interband transitions under variation of carriers concentration by
the gate voltage is studied theoretically. The calculations were performed 
for strongly doped graphene on high-$\kappa$ (Al$_2$O$_3$, HfO$_2$, AlN, 
and ZrO$_2$) or SiO$_2$ substrates under normal propagation of radiation. 
We have obtained the modulation depth above 10\% depending on wavelength,
gate voltage (i.e. carriers concentration), and parameters of substrate.
The graphene - dielectric substrate - doped Si (as gate) structures can be
used as an effective electrooptical modulator of near-IR and mid-IR 
radiation for the cases of high-$\kappa$ and SiO$_2$ substrates, respectively.
\end{abstract}

\pacs{78.67.Wj, 42.79.Hp}

\maketitle

\section{Introduction}
The essential feature of graphene's optical properties is its 
substantial interaction with radiation in the wide spectral region, 
from far-$IR$ up to $UV$, due to effective interband transitions (see 
reviews \cite{1}). The enhancement of this responce due to interference 
permits one to make the graphene on dielectric substrate visible \cite{2}. 
The other example of the exceptional optical properties is the 
graphene-based saturable absorber for ultrafast lasers in the telecommunication 
spectral region \cite{3}. Besides this, the carriers contribution modifies 
essentially the graphene response due to the Pauli blocking effect, when 
absorption is supressed at $\hbar\omega /2< \varepsilon_{F}$, where $\omega$ 
is the frequency of radiation, 
$\varepsilon_{F}$ is the Fermy energy, see experimental data and 
discussion in Refs. 4 and 5. Recently, modulation \cite{5a} and 
polarization \cite{5b} of IR radiation were observed in the graphene 
structure integrated with an optical waveguide. The efficiency of modulation 
can also be enhanced for the case of normal propagation of radiation by the 
interference effect under an appropriate thickness of substrate, see Figs. 
1a and 1b. As a result, an improvement of modulation for transmission 
and reflection coefficients of the graphene - substrate - gate structure 
by a gate voltage takes place in contrast to the case of graphene placed on 
a semi-infinete insulator. \cite{5c} In this paper, we perform the calculation 
of the optical characteristics of such a structure, and we discuss the 
conditions for realisation of the graphene-based modulator in the near-
and mid-IR spectral region.
\begin{figure}[ht]
\begin{center}
\includegraphics{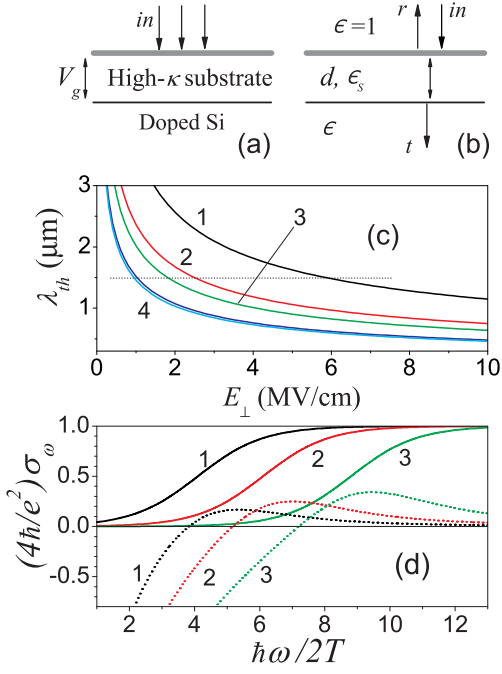}
\end{center}
\addvspace{-0.5 cm}
\caption{(Color online) (a) Schematic view of graphene - high-$\kappa$ substrate - heavily doped Si structure ($V_g$ is gate voltage). (b) Normal propagation of incident ($in$), reflected ($r$), and transmitted ($t$) radiation through the graphene placed over substrate of thickness $d$ on doped Si ($\epsilon_s$ and 
$\epsilon$ are the correspondent dielectric permittivities). (b)
Threshold wavelenthes $\lambda_{th}$ versus transverse field $E_\bot$ 
for structures with different substrates: SiO$_2$ (1), AlN (2), 
Al$_2$O$_3$ (3), and HfO$_2$ (4) (the results for ZrO$_2$ are similar 
to curve (4)). Dashed line is correspondent to the telecommunication 
wavelength, 1.55 $\mu$m. (c) Spectral dependences of Re$\sigma_{\omega}$ 
and Im$\sigma_{\omega}$ (solid and dotted curves) at room temperature 
for concentrations: 10$^{12}$ cm$^{-2}$ (1), $2\times 10^{12}$ cm$^{-2}$ 
(2), $4\times 10^{12}$ cm$^{-2}$ (3).}
\end{figure}

At low temperatures, or high doping levels, the threshold frequency 
for the jump of absorption is determined by the condition 
$\hbar\omega_{th}=2\varepsilon_F\propto\sqrt{n}$, where concentration 
$n$ depends linearly on gate voltage, $V_g$, and is inverse to substrate 
thickness, $d$, so that $n\propto V_g/d\equiv E_{\bot}$. The dependences 
of the threshold wavelength $\lambda_{th}$ on homogeneous field 
$E_{\bot}$ are presented in Fig. 1c for a number of substrates. 
One can see that the modulation of near-IR radiation (1.55 $\mu$m) 
is possible at $E_{\bot}<$3 MV/cm for the high-$\kappa$ substrates 
(Al$_2$O$_3$, HfO$_2$, AlN, and ZrO$_2$ are examined below), and for 
SiO$_2$ substrate the applied field should be twice stonger.
The modulation of mid-IR radiation (10.6 $\mu$m) can be realised in 
lower fieids, $E_{\bot}<$ 1 MV/cm. For the modulation of radiation 
in the visible spectral range, a field $E_{\bot}$ comparable with 
the breakdown field for the substrate under consideration is needed.
The modulation depth can be estimated through the amplitude of the 
absorbtion jump on the threshold, equal to 2.3\%. For the 5 layer 
graphene the modulation efficiency can exceed 10\%, taking into account
that the interference on substrate influences this value. The 
absorption edge spreads with the increase of temperature (see 
the real part of dynamic conductivity $\sigma_\omega$ presented in Fig. 
1d), and the modulation efficiency decreases. At room temperature the 
effective modulation for near-IR radiation only is possible, while the 
modulation of mid-IR radiation needs cooling to temperatures about 77 K.
The efficiency of modulation obtained is comparable to results both in bulk 
semiconductors and in heterostructures, see Refs. 9 and 10, respectively.

The analysis performed below is organized as follows. In Sect. II we 
describe the carrier-induced modulation of the response of graphene and evaluate the transmission and reflection coefficients. The efficiency 
of modulation versus applied field and thickness of substrate is analyzed 
in Sect. III. Discussion and concluding remarks are given in the last
section.

\section{Basic equations}
The response of a doped graphene sheet on probe radiation is described 
by the two-dimensional dynamic conductivity $\sigma_{\omega}$. 
In the colisionless approximation, when $\omega$ exceeds any relaxation 
rate, the real and imaginary parts of $\sigma_{\omega}$ are given
by the expressions:
\begin{eqnarray}  
{\rm Re}\sigma_\omega\simeq\frac{e^2}{4\hbar}( 1-f_{ep_\omega}-
f_{hp_\omega}) , \\
{\rm Im}\sigma_\omega\simeq\bar\sigma_\omega 
-\frac{e^2}{4\hbar p_\omega}{\cal P}\int\limits_0^\infty\frac{dpp^2}
{p_\omega^2 -p^2}\left( f_{ep}+f_{hp}\right) . \nonumber
\end{eqnarray}
In a doped graphene, when $E_{\bot}$ is strong enough, the Pauli blocking factor in ${\rm Re}\sigma_\omega$ is written through electron and hole 
distribution functions, $f_{e,h~p}$, taken at $p_{\omega}=\hbar\omega 
/(2v)$, where $v=10^{8}$ cm/s is the velocity of quasiparticles. Mention, 
that in the absence of carriers ${\rm Re}\sigma_{\omega}$ does not depend 
on any material parameters. \cite{6} In the imagionary part of 
$\sigma_{\omega}$ we eliminate the contribution of the non-doped graphene 
$\bar\sigma_\omega$, see Refs. 12 and 13, and the carriers contribution is connected with ${\rm Re}\sigma_\omega$ through the Kramers-Kronig relation, where $\cal P$ means the principal value of integral. The response of 
$N$-layer epitaxial graphene \cite{9} is described by the total conductivity $N\sigma_{\omega}$.

We restrict ourselvers to the geometry of normal propagation of the 
incident ($in$-), reflected ($r$-), and transient ($t$-) waves through the structure "N-layer graphene - dielectric substrate - doped Si", as it is 
shown in Fig. 1a. The in-plane electric field $E(z)\exp (-i\omega t)$ is governed by the wave equation: \cite{1,8}
\begin{equation} 
\frac{d^2 E(z)}{dz^2}+\epsilon_{\omega}(z)\left(\frac{\omega}{c}
\right)^2E(z)=0 , 
\end{equation}
where $z\neq 0$ and the permittivity $\varepsilon_{\omega}(z)$ is equal 
to the constant $\varepsilon_{s}$ in the substrate layer with the thickness 
$d$ (at $0<z<d$), while in the thick Si the dispersion $\varepsilon_{\omega}$  
should be taken into account, see Ref. 15. The boundary conditions at 
$N$-layer graphene sheet, where $z\rightarrow 0$, takes the form
\begin{equation}
\left.\frac{dE(z)}{dz}\right|_{-0}^{+0}+i\frac{4\pi\omega}{c^2}N\sigma_\omega  
E(z=0)=0, ~~~~ \left. E(z)\right|_{-0}^{+0}=0 .
\end{equation}
These expressions contain the contribution of surface current, proportional to
$N \sigma_{\omega}$, that determines the jump of $[dE(z)/dz]$, while
$E(z)$ is continuous. At the substrate-Si interface we use the two
conditions of continuity:
\begin{equation}
\left.\frac{dE(z)}{dz}\right|_{d-0}^{d+0}=0 , ~~~~
E(z)|_{d-0}^{d+0}=0 .
\end{equation}

Outside of the graphene sheet the solution of electrodynamic problem 
(2)-(4) should be written in the form
\begin{equation}
E(z)=\left\{ \begin{array}{*{20}c}
E_{in}e^{ik_\omega z}+E_r e^{-ik_\omega z} & z<0 \\
E_+ e^{i\tilde k_\omega z}+E_- e^{-i\tilde k_\omega z} & 0<z<d \\
E_t e^{i\bar k_\omega z} & d<z \end{array} \right.  .
\end{equation}
Here the amlitudes for $in$- and $r$-waves ($z<0$), $t$-wave ($z>d$), 
and for the field $E_{\pm}$ in the dielectric substrate (at $0<z<d$) 
are introduced. In Eq.(5) the wave vectors $k_\omega =\omega /c$ (to the left), $\tilde k_\omega =\sqrt{\epsilon_s }\omega /c$ (to the right), 
and $\bar k_\omega =\sqrt{\epsilon_\omega}\omega /c$ (in the dielectric 
substrate) are introduced as well. Using the boundary conditions
(3) and (4) we get the system of linear equations for the amlitudes
above. The solution of such a system determines the transition 
and reflection coefficients, $T_\omega$ and $R_\omega$, according to
\begin{equation}
T_\omega =\sqrt{\epsilon_\omega}\frac{|E_t|^2}{E_{in}^2},
 ~~~~~  R_\omega =\frac{|E_r|^2}{E_{in}^2} .
\end{equation}
According to energy conservation law, which connects $T_\omega$ and $R_\omega$
with the relative absorption coefficient $\xi_\omega$, one obtains:
\begin{equation}
T_\omega +R_\omega +\xi_\omega =1 .
\end{equation}
As a result, variations of $T_\omega$ and $R_\omega$ are correlated due  
to the Pauli blocking effect which leads to a jump of $\xi_\omega$.

The direct expressions for the transmission and reflection
coefficients take form 
\begin{equation}
T_\omega =\frac{4\sqrt{\epsilon_\omega}}{|{A_\omega^{(+)}}|^2}, ~~~~~~
R_\omega =\left|\frac{A_\omega^{(-)}}{A_\omega^{(+)}}\right|^2 .
\end{equation}
where $A^{(\pm )}_{\omega}$ are expressed through the dynamic conductivity
and the structure parameters according to
\begin{eqnarray}
A_\omega^{(\pm )}=\sqrt{\epsilon}\cos\tilde k_\omega d-i\sqrt{\epsilon_s} 
\sin\tilde k_\omega  d  \\
+\left(\frac{4\pi\sigma_\omega}{c}\pm 1\right)\left( {\cos\tilde k_\omega
d-i\sqrt{\varepsilon /\varepsilon_s}\sin\tilde k_\omega d} \right) . 
\nonumber
\end{eqnarray}
In the case $\epsilon_\omega =\epsilon_s$ the oscillating factors eliminate 
from (9), and Eqs. (8) are in agreement with the previous results. \cite{1,8} 
Taking into consideration the interference on the substrate, when 
$\epsilon_\omega\neq\epsilon_s$, the spectral dependences of $T_{\omega}$ 
and $R_{\omega}$ are determined by carriers concentration (through 
variations of $E_{\bot}$ or $V_g$), the dielectric substrate thickness 
$d$, and the permittivities $\varepsilon_s$ and $\varepsilon_{\omega}$. 
We have neglected a weak absorption in Si and used $\varepsilon_s$ for 
SiO$_2$ and high-$\kappa$ dielectrics from Refs. 15 and 16, respectively.

\section{Results}
Performing the numerical integration in Eq. (1) and using Eqs. (8, 9) 
one obtains the transmission and reflection coefficients. Below we 
analyze the dependences of $T_\omega$ and $R_\omega$ on the applied 
field $E_{\bot}$, which determines the carriers concentration, for the substrates of various thickness on the base of high-$\kappa$ dielectrics 
or SiO$_2$. The computations were performed for near-IR and mid-IR 
spectral regions ($\lambda =$1.55 $\mu$m and 10.6 $\mu$m).
\begin{figure}[ht]
\begin{center}
\includegraphics{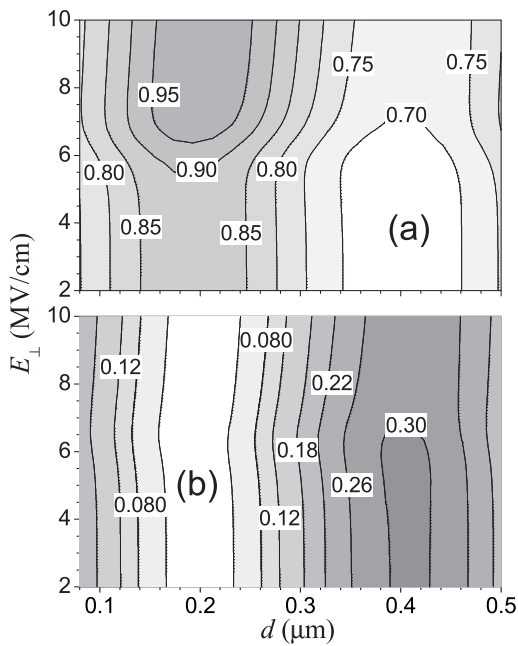}
\end{center}
\addvspace{-0.5 cm}
\caption{(Color online) (a) Contour plot of transmission coefficient of 
graphene over Al$_2$O$_3$ substrate as a function of applied field $E_{\bot}$ and thickness $d$. (b) The same for reflection coefficient. } 
\end{figure}

\subsection{High-$\kappa$ substrates}
We examine first the modulation of the telecommunication range radiation, 
$\lambda =$1.55 $\mu$m, by the structures of 5-layer graphene on 
high-$\kappa$ substrate at room temperature. Figures 2(a) and 2(b) show the contour plots 
of $T$ and $R$ versus $E_{\bot}$ and $d$ for the case of Al$_2$O$_3$ 
substrate. One can see, that the change of $T$ versus $E_{\bot}$ is 
$\sim$10\% near the transmission maximum, and the modification of $T$ 
by interference can be as large as 0.3 if $d$ is in the range 0.1 - 0.5 $\mu$m. 
Similarly, the change of $R$ versus $E_{\bot}$ does not exceed 
several \%, while the modification of $R$ versus $d$ can be of 0.3 order. 
The maximum of $T$ corresponds the minimum of $R$ and vise versa.
\begin{figure}[ht]
\begin{center}
\includegraphics{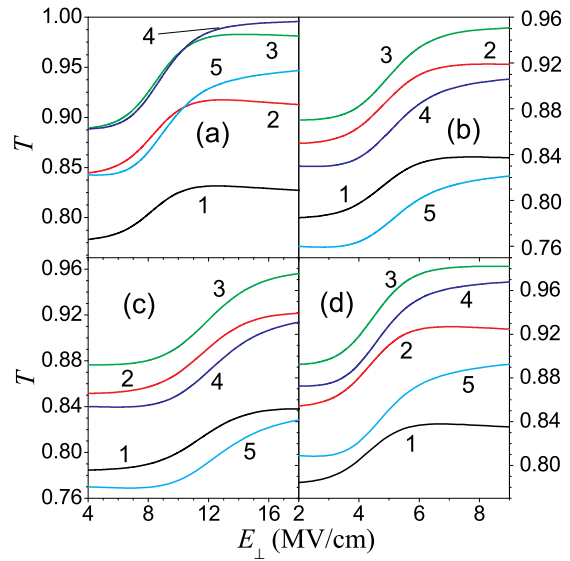}
\end{center}
\addvspace{-0.5 cm}
\caption{(Color online) Transmissivity $T$ at wavelength 1.55 $\mu$m versus 
$E_{\bot}$ for different substrates: (a) Al$_2$O$_3$, (b) HfO$_2$, (c) AlN, 
and (d) ZrO$_2$. Curves 1 - 5 are correspondent to the thicknesses $d=$0.08,
0.12, 0.16, 0.2, and 0.24 $\mu$m. }
\end{figure}
\begin{figure}[ht]
\begin{center}
\includegraphics{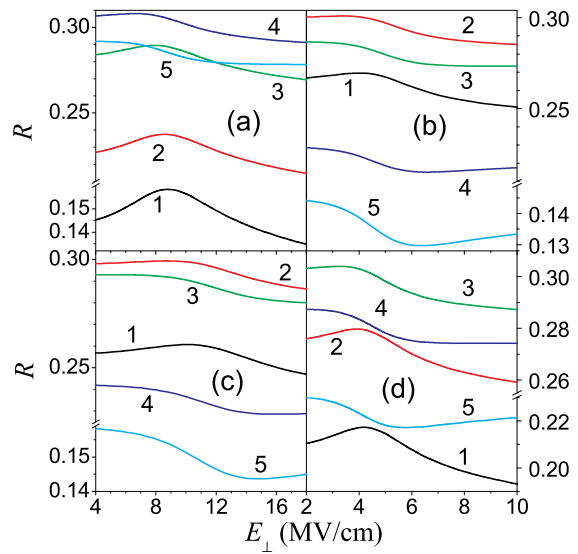}
\end{center}
\addvspace{-0.5 cm}
\caption{(Color online) Reflectivity $R$ at wavelength 1.55 $\mu$m versus 
$E_{\bot}$ for the same substrates as in Fig. 3(a-d). Curves 1 - 5 are correspondent to the thicknesses $d=$0.28, 0.32, 0.36, 0.4 and 0.44 $\mu$m. }
\end{figure}

Gate-voltage-induced modification of transmission is presented in Fig. 3 
for different high-$\kappa$ substrates at several thicknesses near the 
maximum of $T$. Similar dependences for $R$ are presented in Fig. 4 
near the reflection maximum, corresponding to the greater thicknesses. 
Besides the essential dependence on thickness, $T$ and $R$ depend also 
on high-frequency and static permittivity of the materials under 
consideration. Therefore, the effective modulation for
Al$_2$O$_3$ and AlN occurs in the range $E_{\bot}\sim$8 - 12 MV/cm,
and for HfO$_2$ and ZrO$_2$ the weaker fields $E_{\bot}\sim$4 - 6 MV/cm 
are needed. The modulation depth for transmission 
exceeds in several times the modulation depth for 
reflection. The effective modulation interval, corresponding
the region of the jump in absorption, becomes narrower
with the decrease of temperature.

\subsection{SiO$_2$ substrate}
Now we are going to examine the structures "graphene - SiO$_2$ - Si", where 
the permittivities are smaller. Therefore the effective modulation for 
transmission of near-IR radiation takes place at $E_{\bot}\sim$25 - 35 
MV/cm, see Fig. 5(a), i.e. it needs stronger fields than the
threshold of single-layer graphene on SiO$_2$ substrate $\sim$6 MV/cm, 
see Fig. 1(b). The modulation of $R$ in the same range of fields does
not exceed several percents, see Fig. 5(b). It should be noted, that these 
fields $E_{\bot}$ are of the same order of values, as a breakdown field, 
therefore a possibility of modulation in this case needs a special 
verification.
\begin{figure}[ht]
\begin{center}
\includegraphics{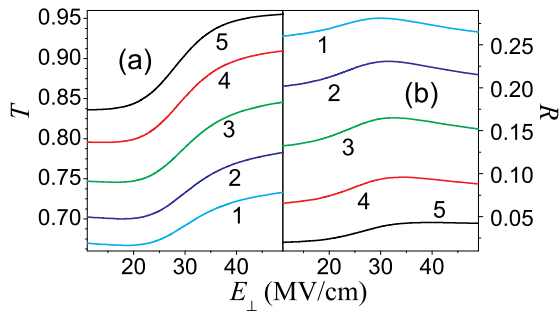}
\end{center}
\addvspace{-0.5 cm}
\caption{(Color online) Transmissivity (a) and reflectivity (b), $T$ and
$R$, at wavelength 1.55 $\mu$m versus $E_{\bot}$ for SiO$_2$  substrate. 
Curves 1 - 5 are correspondent to the thicknesses $d=$0.24, 0.28, 0.32, 
0.36 and 0.4 $\mu$m.}
\end{figure}

The effective modulation of transmission (over 5\%, see Fig. 6(a)) 
takes place in mid-IR spectral region, for $\lambda =$10.6 $\mu$m. 
The applied field in this case does not exceed 2 MV/cm, but
the substrate thickness should be greater because of
the increase of $\lambda$. The corresponding modulation
of reflection does not exceed several percents as well,
see Fig. 6(b).
\begin{figure}[ht]
\begin{center}
\includegraphics{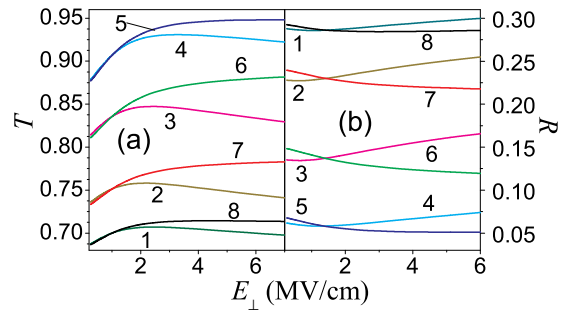}
\end{center}
\addvspace{-0.5 cm}
\caption{(Color online) Transmissivity (a) and reflectivity (b) versus 
$E_{\bot}$ for five-layer graphene over SiO$_2$  substrate at wavelength 
10.6 $\mu$m and room temperature. Curves 1 - 8 are correspondent to the thicknesses $d=$0.3, 0.6, 0.9, 1.2, 1.5, 1.8, 2.1 and 2.4 $\mu$m.}
\end{figure}

For the single layer graphene the modulation depth
obviously can not be greater, than 2.3\%. However, this
modulation of transmission occurs for much lower fields, than 
in the previous cases under examination, $~$200 kV/cm,
and the jump region becomes rather narrow with the
decrease of temperature, see Figs. 7a, b.

\begin{figure}[ht]
\begin{center}
\includegraphics{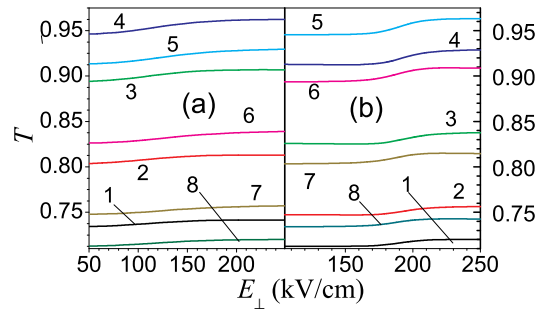}
\end{center}
\addvspace{-0.5 cm}
\caption{(Color online) Transmissivity at wavelength 10.6 $\mu$m versus 
$E_{\bot}$ for a sigle layer graphene over SiO$_2$  substrate at temperatures 77 K 
(a) and 20 K (b). Curves 1 - 8 are correspondent to the same thicknesses 
as in Fig. 6.}
\end{figure}

\section{Concluding remarks}

The results obtained clearly demonstrate the possibility for 
realization of the  modulator for telecommunication spectral range 
($\sim$1.5 $\mu$m) on the base of multilayer graphene, placed over 
high-$\kappa$ substrate. The effective modulation can be realized in 
this case for the applied fields $\sim$5 MV/cm, while for the case of
SiO$_2$ substrate the field should be $\sim$20 MV/cm comparable to
a breakdown value. The modulation depth for multi-layer ($N=$5-10)
graphene can be as large as 10-20\% ($\sim$2\% per one layer).
The same efficiency of modulation for the mid-IR radiation
($\sim$10.6 $\mu$m) can be realized for the applied fields 
not stronger than 2 MV/cm for the low temperature region. 

The consideration performed takes into account the contribution of 
interband transitions, described by the complex dynamic conductivity, 
and the radiation interference on the structure "vacuum - graphene - 
substrate - doped Si" for the case of normal propagation of radiation.
The modulation is determined by the Pauli blocking effect under the change 
of the carriers concentration by the gate voltage, therefore the time 
of modulation is governed by the recombination time of the excess  
concentration of carriers, or by the time of injection from contacts.

Next, we discuss the assumptions used in our calculations, which are 
rather standart ones. The dynamic conductivity of the carriers in the 
spectral region under examination is described properly with the use of
the linear dispersion law of carriers in graphene. The phenomenological 
description of the dispersion of $Im\sigma_\omega$ due to the transitions
from the valence band (see Refs. 12 and 13) does not change the results
essentially due to the smallness of its contribution for the spectral 
range under consideration. The study of the declined propagation of 
radiation is more complicated, and the modulation efficiency in this 
case decreases. Moreover, the modulation efficiency can also be reduced 
in mid-IR range due to the absorption of radiation in the doped Si. It 
should be noted, that modulation of electron concentration in the gate 
gives a weak contribution to IR response and can be neglected in 
comparison to the Pauli blocking effect in graphene.

In conclusion, the results obtained should stimulate the experimental study of 
the electrooptical modulation of the near-IR radiation by the structure of 
multilayer graphene over high-$\kappa$ dielectric substrate at room temperature 
and high gate voltages (concentrations). For the mid-IR spectral region the 
effective modulation can be realized at low temperatures.  

\begin{acknowledgments}
This work was supported by the State Fundamental Research Fund of Ukraine (Grant 40.2/069).
\end{acknowledgments}

\end{document}